\documentclass[aps,prc,preprint,showpacs,showkeys]{revtex4}
\usepackage{amssymb}
\usepackage{amsmath}
\usepackage{graphicx}
\usepackage{bm}
\usepackage{epsf}

\begin{document}

\title{Neutron star matter in the quark-meson coupling model in strong
magnetic fields}
\author{P. Yue}
\email{yuep@mail.nankai.edu.cn}
\affiliation{Department of Physics, Nankai University, Tianjin 300071, China}
\author{H. Shen}
\email{songtc@nankai.edu.cn}
\affiliation{Department of Physics, Nankai University, Tianjin 300071, China}

\begin{abstract}
The effects of strong magnetic fields on neutron star matter are investigated
in the quark-meson coupling (QMC) model. The QMC model describes a nuclear
many-body system as nonoverlapping MIT bags in which quarks interact through
self-consistent exchange of scalar and vector mesons in the mean-field approximation.
The results of the QMC model are compared with those obtained in a relativistic
mean-field (RMF) model. It is found that quantitative differences exist between
the QMC and RMF models, while qualitative trends of the magnetic field effects
on the equation of state and composition of neutron star matter are very similar.
\end{abstract}

\maketitle


\section{Introduction}

\label{introduction}

Recent observations of soft gamma repeaters (SGRs) and anomalous X-ray
pulsars (AXPs) have suggested that the surface magnetic field of young
neutron stars could be of order $10^{14}-10^{15}$ G~\cite{TD95}. On the other
hand, it is estimated that the interior field in neutron stars may be as large
as $10^{18}$ G~\cite{LS91,Latt00}. Motivated by the existence of strong
magnetic fields in neutron stars, theoretical studies on the effect of
extremely large fields on dense matter and neutron stars have been carried
out by many authors~\cite{Latt00,Latt01,CBP97,Yuan99,Mao01,Mao03,Mao06,prd02}.
The inclusion of hyperons and boson condensation has
also been investigated~\cite{Latt02,Pion01,Kaon02}.
It was found that magnetic fields could change the
composition of matter dramatically, and the softening of the equation of
state (EOS) caused by Landau quantization is overwhelmed by the stiffening due
to the incorporation of the nucleon anomalous magnetic moments for field strengths
$B>10^{5}\, B_c^e$ ($B_c^e=4.414\times 10^{13}$ G is the electron critical
field). So far, most of these studies have employed the relativistic mean-field (RMF)
theory, which is a field theoretical approach at hadron level and has been
widely used in the description of nuclear matter and finite nuclei~\cite{WS86}.
It is well known that various RMF models could provide similar nuclear saturation
properties but different behaviors at high density. In Ref.~\cite{Latt00},
the model dependence of strong magnetic field effects within the RMF
theory has been investigated. It is instructive to study the properties of
neutron star matter in the presence of strong magnetic fields by employing
various approaches.

In this paper, we adopt the quark-meson coupling (QMC) model for the
investigation of strong magnetic field effects on neutron star matter.
The QMC model was originally proposed by Guichon~\cite{QMC88},
in which the quark degrees of freedom are explicitly taken into account.
The QMC model describes a nuclear
many-body system as nonoverlapping MIT bags in which quarks interact
through self-consistent exchange of scalar and vector mesons in the
mean-field approximation. The mesons couple not to point-like nucleons but
directly to confined quarks. In contrast to the RMF approach, the quark
structure of nucleons plays a crucial role in the QMC model, and the basic
coupling constants are defined at quark level. The QMC model has been
subsequently extended and applied to various problems of nuclear matter and
finite nuclei with reasonable success~\cite{QMC96,QMC98,QMC98h}.
Furthermore, the model has also been used to investigate the properties
of neutron stars with the inclusion of hyperons, quarks, and kaon
condensation~\cite{prc99,Panda04,Panda05}. It is interesting to extend the QMC
model for the study of a nuclear many-body system in strong magnetic fields,
and make a systematic comparison between the QMC and RMF models.

In general, a nuclear many-body problem has to be worked out in the QMC
model by two steps. First, one can calculate the nucleon properties in
medium by using the MIT bag model with an external scalar field $\sigma$
generated by nuclear medium. Then in the second step, the entire nuclear
system is solved at hadron level with the effective nucleon mass as a
function of $\sigma$ obtained in the first step. The main difference between
the QMC and RMF models is reflected in the dependence of effective nucleon
mass on scalar field $\sigma$. In the QMC model, the scalar field plays a
vital role in determining the nucleon properties such as effective mass and
radius, while the vector fields do not cause any change of nucleon
properties but appear merely as energy shifts. When we treat a nuclear
many-body system in strong magnetic fields, we assume that the magnetic
fields do not change the internal structure of nucleons, then nucleons in
meson and magnetic fields could be treated quite similarly as in the RMF
models. With the effective nucleon mass obtained at quark level, we can
investigate the properties of neutron star matter in strong magnetic fields.

This paper is arranged as follows. In Sec. II, we briefly describe the QMC
model for neutron star matter and present relevant equations in strong
magnetic fields. In Sec. III, we show and discuss the numerical results in
the QMC model and make a systematic comparison with the results of the RMF
model. The conclusion is presented in Sec. IV.


\section{ Quark-Meson Coupling model for neutron star matter in strong
magnetic fields}

The QMC model describes a nuclear many-body system as nonoverlapping
spherical bags in which quarks interact through self-consistent exchange
of scalar and vector mesons in the mean-field approximation. To perform the
calculation for neutron star matter in strong magnetic fields, we first
study the nucleon properties with external meson and magnetic fields by
using the MIT bag model. The fields in the bag are in principle functions
of position, which may cause a deformation of the nucleon bag. For
simplicity, we neglect the spatial variation of the fields over the small
bag volume, and take the values at the center of the bag as their average
quantities~\cite{QMC96}. Then the quarks in the bag satisfy the Dirac equation
\begin{equation}
\left[ i\gamma _{\mu }\partial ^{\mu }-\left( m_{q}+g_{\sigma }^{q}\sigma
\right) -g_{\omega }^{q}\omega _{\mu }\gamma ^{\mu }-g_{\rho }^{q}\tau
_{3}\rho _{3\mu }\gamma ^{\mu }-\frac{e\left( 1+3\tau _{3}\right) }{6}A_{\mu
}\gamma ^{\mu }\right] \psi_{q} = 0 \ ,
\end{equation}%
where $g_{\sigma }^{q}$, $g_{\omega }^{q}$, and $g_{\rho }^{q}$ are the
quark-meson coupling constants, and $m_{q}$ is the current quark mass. $%
\sigma $, $\omega _{\mu }$, $\rho _{3\mu }$, and $A_{\mu }$\ are the values
of the meson and magnetic fields at the center of the nucleon bag. The
normalized ground state for a quark in the bag is given by%
\begin{equation}
\psi_{q}(\mathbf{r},t)=\mathcal{N}_q\,e^{-i\epsilon_{q}t/R}\left(
\begin{array}{c}
j_{0}(x_q r/R) \\
i\,\beta_{q}\,\vec{\sigma} \cdot \hat{\mathbf{r}}\,j_{1}(x_q r/R)%
\end{array}%
\right) \,{\frac{\chi_{q}}{\sqrt{4\pi }}}\ ,
\end{equation}%
where
\begin{equation}
\beta_{q}=\sqrt{{\frac{\Omega _{q}-R\,m_{q}^{\ast }}
                      {\Omega _{q}\,+R\,m_{q}^{\ast }}}}\ ,
\end{equation}%
\begin{equation}
\mathcal{N}_q^{-2}=2\,R^{3}\,j_{0}^{2}(x_q)\left[ \Omega _{q}
(\Omega_{q}-1)+R\,m_{q}^{\ast }/2\right] /x_q^{2}\ ,
\end{equation}%
with $\Omega_{q}=\sqrt{x_q^{2}+(R\,m_{q}^{\ast })^{2}}$,
$m_{q}^{\ast}=m_{q}+g_{\sigma }^{q}\sigma$. $R$ is the bag radius,
and $\chi_{q}$ is the quark spinor. The boundary condition,
$j_{0}(x_{q})=\beta _{q}\,j_{1}(x_{q})$, at
the bag surface determines the eigenvalue $x_{q}$. The energy of a static
nucleon bag consisting of three ground state quarks is then given by
\begin{equation}
E_{\mathrm{bag}}=3{\frac{\Omega _{q}}{R}}-{\frac{Z}{R}}+{\frac{4}{3}}\pi
R^{3}B_{bag}\ ,
\end{equation}%
where the parameter $Z$ accounts for zero-point motion, and $B_{\textrm{bag}}$
is the bag constant. The effective nucleon mass is then taken to be
\begin{equation}
M_{N}^{\ast }=E_{\textrm{bag}}\ .
\end{equation}%
The bag radius $R$ is determined by the equilibrium condition $\partial
M_{N}^{\ast }/\partial R=0$. In the present calculation, we take the current
quark mass $m_{q}=5.5$ MeV. The parameter $B_{\textrm{bag}}^{1/4}=210.854$ MeV
and $Z=4.00506$ are determined by reproducing the nucleon mass in free space
$M_{N}=939$ MeV and the bag radius $R=0.6$ fm as given in Ref.~\cite{Panda04}.
We note that $\omega$ and $\rho$ mean fields appear merely as the energy shifts
which do not cause any change in nucleon properties, therefore the effective
nucleon mass obtained in the QMC model depends on the $\sigma $ mean field
only.

For neutron star matter consisting of a neutral mixture of neutrons,
protons, electrons, and muons in $\beta $ equilibrium in the presence of
strong magnetic fields, the total Lagrangian density at hadron level can be
written as%
\begin{eqnarray}
\label{eq:lqmc}
\mathcal{L_{\mathrm{QMC}}} &=&\sum_{b=n,p}\bar{\psi}_{b}\left[ i\gamma _{\mu
}\partial ^{\mu }-q_{b}\gamma _{\mu }A^{\mu }-M_{N}^{\ast }-g_{\omega
}\gamma _{\mu }\omega ^{\mu }-g_{\rho }\gamma _{\mu }\tau _{i}\rho _{i}^{\mu
}-\frac{1}{2}\kappa _{b}\sigma _{\mu \nu }F^{\mu \nu }\right] \psi _{b}
\notag \nonumber \\
&&+\sum_{l=e,\mu }\bar{\psi_{l}}\left[ i\gamma _{\mu }\partial ^{\mu
}-q_{l}\gamma _{\mu }A^{\mu }-m_{l}\right] \psi_{l}+\frac{1}{2}\partial
_{\mu }\sigma \partial ^{\mu }\sigma -\frac{1}{2}m_{\sigma }^{2}\sigma ^{2}
\nonumber \\
&&-\frac{1}{4}W_{\mu \nu }W^{\mu \nu }+\frac{1}{2}m_{\omega }^{2}\omega
_{\mu }\omega ^{\mu }-\frac{1}{4}R_{i\mu \nu }R_{i}^{\mu \nu }+%
\frac{1}{2}m_{\rho }^{2}\rho _{i\mu }\rho _{i}^{\mu }-\frac{1}{4}%
F_{\mu \nu }F^{\mu \nu } \ ,
\end{eqnarray}%
where $\psi _{b}$ and $\psi _{l}$ are the nucleon and lepton fields,
respectively. $A^{\mu }= (0$, $0$, $Bx$, $0)$ refers to a constant
external magnetic field $B$ along the $z$-axis. The field tensors for the $%
\omega $, $\rho $, and magnetic field are given by $W_{\mu \nu }=\partial
_{\mu }\omega _{\nu }-\partial _{\nu }\omega _{\mu }$, $R_{i\mu \nu
}=\partial _{\mu }\rho _{i\nu }-\partial _{\nu }\rho _{i\mu }$, and $F_{\mu
\nu }=\partial _{\mu }A_{\nu }-\partial _{\nu }A_{\mu }$.
It has been argued in Ref.~\cite{Latt00} that the contributions from anomalous
magnetic moments of nucleons should be considered for large field strengths.
Here the anomalous magnetic moments of nucleons are included with
$\kappa _{p}=\mu_{N}\left(g_{p}/2-1\right) =1.7928\ \mu _{N}$ and
$\kappa _{n}=\mu_{N}g_{n}/2=-1.9130\ \mu _{N}$, where $\mu _{N}$ is the nuclear magneton.
As for leptons, the effects of anomalous magnetic moments
of electrons and muons on the EOS are very small as shown in Ref.~\cite{Mao06},
and the electron anomalous magnetic moments could be efficiently reduced by high-order
contributions from the vacuum polarization in strong magnetic fields~\cite{eAMM}.
Therefore, we neglect the anomalous magnetic moments of leptons
in the present work.
The coupling constants at hadron level, $g_{\omega }$ and $g_{\rho }$, are
related to the corresponding quark-meson couplings as $g_{\omega
}=3g_{\omega }^{q}$ and $g_{\rho }=g_{\rho }^{q}$~\cite{QMF00}. The quark-meson coupling
constants $g_{\sigma }^{q}=5.957$, $g_{\omega }^{q}=2.994$, and $g_{\rho
}^{q}=4.325$ are determined by fitting the properties of nuclear matter~\cite{Panda04},
and the meson masses $m_{\sigma }=550$ MeV, $m_{\omega }=783$ MeV, and $m_{\rho
}=770$ MeV are used in the present calculation.

From the Lagrangian density given in Eq.~(\ref{eq:lqmc}), we obtain the following meson
field equations in the mean-field approximation
\begin{eqnarray}
\label{eq:m1}
m_{\sigma }^{2}\sigma &=&-\frac{\partial {M_{N}^{\ast }}}{\partial \sigma }%
(\rho _{s}^{p}+\rho _{s}^{n}) \ ,
\\ \label{eq:m2}
m_{\omega }^{2}\omega _{0} &=&g_{\omega }(\rho _{v}^{p}+\rho _{v}^{n}) \ ,
\\ \label{eq:m3}
m_{\rho }^{2}{\rho _{30}} &=&g_{\rho }(\rho _{v}^{p}-\rho _{v}^{n}) \ ,
\end{eqnarray}%
while the Dirac equations for nucleons and leptons are given by%
\begin{eqnarray}
 & \left( {i\gamma _{\mu }\partial ^{\mu }-q_{b}\gamma _{\mu }A^{\mu
}-M_{N}^{\ast }-g_{\omega }\gamma ^{0}\omega _{0}-g_{\rho }\gamma ^{0}\tau_3\rho
_{30}-\frac{1}{2}\kappa _{b}\sigma _{\mu \nu }F^{\mu \nu }}\right) \psi
_{b} = 0 \ , \\
 & \left( {i\gamma _{\mu }\partial ^{\mu }-q_{l}\gamma _{\mu }A^{\mu }-m_{l}}%
\right) \psi _{l}= 0 \ .
\end{eqnarray}%
The energy spectra for protons, neutrons, and leptons (electrons and muons) are given by%
\begin{eqnarray}
E_{\nu,s}^{p} &=&\sqrt{k_{z}^{2}+\left( \sqrt{{M_{N}^{\ast 2}+}2\nu {q_{p}B}%
}-s{\kappa _{p}B}\right) ^{2}}+g_{\omega }\omega _{0}+g_{\rho }{\rho _{30}} \ ,
\\
E_{s}^{n} &=&\sqrt{k_{z}^{2}+\left( \sqrt{{M_{N}^{\ast 2}}
+k_{x}^{2}+k_{y}^{2}}-s{\kappa_{n}B}\right) ^{2}}+g_{\omega }\omega
_{0}-g_{\rho }{\rho _{30}} \ , \\
E_{\nu ,s}^{l} &=&\sqrt{k_{z}^{2}+m_{l}^{2}{+}2\nu \left\vert {q_{l}}%
\right\vert {B}} \ ,
\end{eqnarray}%
where $\nu =n+1/2-sgn\left(q\right)s/2=0,1,2,\ldots $ enumerates
the Landau levels of the fermion with electric charge $q$. The quantum
number $s=\pm 1$ are for spin-up and spin-down cases. As discussed in
Ref.~\cite{Latt00}, the Dirac spinors are no longer eigenfunctions of
the spin operator along the magnetic field direction $\sigma_{z}$ when the anomalous
magnetic moments are taken into account. However, $s$ reduce to the eigenvalue
of $\sigma _{z}$ as the anomalous magnetic moment tends toward zero.
We note that the first Landau level to be occupied is the state with $\nu =0,$ $s=+1$
for the proton and $\nu =0,$ $s=-1$ for the lepton, while the neutron first occupies
the state with $s=-1$ due to its negative anomalous magnetic moment.
This argument is consistent with the results in nonrelativistic approach~\cite{Latt00}.

The expressions of the scalar and vector densities for protons and neutrons
are given by%
\begin{eqnarray}
\rho _{s}^{p} &=&\frac{{q_{p}B}M_{N}^{\ast }}{2\pi ^{2}}\sum_{\nu }\sum_{s}%
\frac{\sqrt{M_{N}^{\ast 2}+2\nu {q_{p}B}}-s\kappa _{p}B}{\sqrt{M_{N}^{\ast
2}+2\nu {q_{p}B}}}\ln \left\vert \frac{k_{f,\nu ,s}^{p}+E_{f}^{p}}{\sqrt{%
M_{N}^{\ast 2}+2\nu {q_{p}B}}-s\kappa _{p}B}\right\vert \ , \\
\rho _{v}^{p} &=&\frac{{q_{p}}B}{2\pi ^{2}}\sum_{\nu }\sum_{s}k_{f,\nu
,s}^{p}\ , \\
\rho _{s}^{n} &=&\frac{M_{N}^{\ast }}{4\pi ^{2}}\sum_{s}\left[
k_{f,s}^{n}E_{f}^{n}-\left( M_{N}^{\ast }-s{\kappa _{n}B}\right) ^{2}\ln
\left\vert \frac{k_{f,s}^{n}+E_{f}^{n}}{M_{N}^{\ast }-s{\kappa _{n}B}}%
\right\vert \right] \ , \\
\rho _{v}^{n} &=& \frac{1}{2\pi ^{2}}\sum_{s}\left\{ \frac{1}{3}k_{f,s}^{n3}-%
\frac{1}{2}s\kappa _{n}B\left[ \left( M_{N}^{\ast }-s{\kappa _{n}B}\right)
k_{f,s}^{n} \right. \right. \nonumber \\
 & & \left. \left. +E_{f}^{n2}\left( \arcsin \frac{M_{N}^{\ast }-s{\kappa _{n}B}}
{E_{f}^{n}}-\frac{\pi }{2}\right) \right] \right\} \ ,
\end{eqnarray}%
where $k_{f,\nu ,s}^{p}$ and $k_{f,s}^{n}$ are the Fermi momenta of protons
and neutrons, which are related to the Fermi energies
$E_{f}^{p}$ and $E_{f}^{n}$ as
\begin{eqnarray}
{E_{f}^{p}}^2 &=&k_{f,\nu ,s}^{p2}+\left( \sqrt{M_{N}^{\ast 2}+2\nu {q_{p}B}}%
-s\kappa _{p}B\right) ^{2}\ , \\
{E_{f}^{n}}^2 &=&k_{f,s}^{n2}+\left( M_{N}^{\ast }-s\kappa _{n}B\right) ^{2}\ .
\end{eqnarray}%
The chemical potentials of nucleons and leptons are given by%
\begin{eqnarray}
\mu _{p} &=&E_{f}^{p}+g_{\omega }\omega _{0}+g_{\rho }{\rho _{30}}\ , \\
\mu _{n} &=&E_{f}^{n}+g_{\omega }\omega _{0}-g_{\rho }{\rho _{30}}\ , \\
\mu _{l} &=&E_{f}^{l} = \sqrt{k_{f,\nu,s}^{l 2} + m_{l}^{2}
                        + 2\nu\left\vert{q_{l}}\right\vert B}\ .
\end{eqnarray}%
For neutron star matter with uniform distributions, the composition of matter is
determined by the requirements of charge neutrality and $\beta $-equilibrium
conditions. In the present calculation, the $\beta $-equilibrium conditions
are expressed by
\begin{eqnarray}
\label{eq:beta1}
\mu _{p} &=&\mu _{n}-\mu _{e} \ ,  \\
\mu _{\mu } &=&\mu _{e} \ ,
\label{eq:beta2}
\end{eqnarray}
and the charge neutrality condition is given by%
\begin{equation}
\rho _{v}^{p}=\rho _{v}^{e}+\rho _{v}^{\mu } \ ,
\label{eq:charge}
\end{equation}%
\bigskip where the vector density of leptons has a similar expression to
that of protons
\begin{equation}
\rho _{v}^{l}=\frac{\left\vert {q_{l}}\right\vert B}{2\pi ^{2}}\sum_{\nu
}\sum_{s}k_{f,\nu ,s}^{l} \ .
\end{equation}%
We solve the coupled Eqs.~(\ref{eq:m1})-(\ref{eq:m3}), (\ref{eq:beta1}),
(\ref{eq:beta2}), and (\ref{eq:charge}) self-consistently at a given baryon density
$\rho_{B}=\rho _{v}^{p}+\rho _{v}^{n}$ in the presence of strong magnetic fields.
The energy density of neutron star matter is given by
\begin{equation}
\varepsilon_m =\varepsilon _{p}+\varepsilon _{n}+\varepsilon _{e}+\varepsilon
_{\mu }+\frac{1}{2}m_{\sigma }^{2}\sigma ^{2}+\frac{1}{2}m_{\omega
}^{2}\omega _{0}^{2}+\frac{1}{2}m_{\rho }^{2}{\rho _{30}^{2}} \ ,
\end{equation}%
where the energy densities of nucleons and leptons have the following forms:%
\begin{eqnarray}
\varepsilon _{p} &=&\frac{{q_{p}B}}{4\pi ^{2}}\sum_{\nu }\sum_{s}\left[
k_{f,\nu ,s}^{p}E_{f}^{p}+\left( \sqrt{M_{N}^{\ast 2}+2\nu {q_{p}B}}-s\kappa
_{p}B\right) ^{2}\ln \left\vert \frac{k_{f,\nu ,s}^{p}+E_{f}^{p}}{\sqrt{%
M_{N}^{\ast 2}+2\nu {q_{p}B}}-s\kappa _{p}B}\right\vert \right] \ , \\
\varepsilon _{n} &=&\frac{1}{4\pi ^{2}}\sum_{s}\left\{ \frac{1}{2}%
k_{f,s}^{n}E_{f}^{n3}-\frac{2}{3}s\kappa _{n}BE_{f}^{n3}\left( \arcsin \frac{%
M_{N}^{\ast }-s{\kappa _{n}B}}{E_{f}^{n}}-\frac{\pi }{2}\right) -\left(
\frac{s{\kappa _{n}B}}{3}+\frac{M_{N}^{\ast }-s{\kappa _{n}B}}{4}\right)
\right.  \notag \\
&&\times \left. \left[ \left( M_{N}^{\ast }-s{\kappa _{n}B}\right)
k_{f,s}^{n}E_{f}^{n}+\left( M_{N}^{\ast }-s{\kappa _{n}B}\right) ^{3}\ln
\left\vert \frac{k_{f,s}^{n}+E_{f}^{n}}{M_{N}^{\ast }-s{\kappa _{n}B}}%
\right\vert \right] \right\} \  , \\
\varepsilon _{l} &=&\frac{\left\vert {q_{l}}\right\vert B}{4\pi ^{2}}%
\sum_{\nu }\sum_{s}\left[ k_{f,\nu ,s}^{l}E_{f}^{l}+\left( m_{l}^{2}{+}2\nu
\left\vert {q_{l}}\right\vert {B}\right) \ln \left\vert \frac{k_{f,\nu
,s}^{l}+E_{f}^{l}}{\sqrt{m_{l}^{2}{+}2\nu\left\vert{q_{l}}\right\vert{B}}}%
\right\vert \right] \  .
\end{eqnarray}%
The pressure of the system can be obtained by
\begin{equation}
P_m=\sum_{i}\mu _{i}\rho _{v}^{i}-\varepsilon_m =\mu _{n}\rho _{B}-\varepsilon_m ,
\end{equation}%
where the charge neutrality and $\beta $-equilibrium conditions are used to
get the last equality.  Note that the contribution from electromagnetic
fields to the energy density and pressure, $\varepsilon_f=P_f=B^2/8\pi$,
should be taken into account in the calculation of the EOS.


\section{ Results and discussion}

In this section, we present the properties of neutron star matter in strong
magnetic fields by using the QMC model. The effective nucleon masses in the QMC
model are obtained self-consistently at quark level, which is the main
difference from the RMF model. The comparison of results in several RMF
models has been investigated extensively in Ref.~\cite{Latt00}. It was found
that quantitative differences persist between those models, while qualitative
trends of magnetic field effects are very similar.
Here we would like to make a systematic comparison between the QMC and RMF models,
which are based on different degrees of freedom.

\begin{figure}[htb]
\includegraphics[bb=30 280 535 755, width=12 cm,clip]{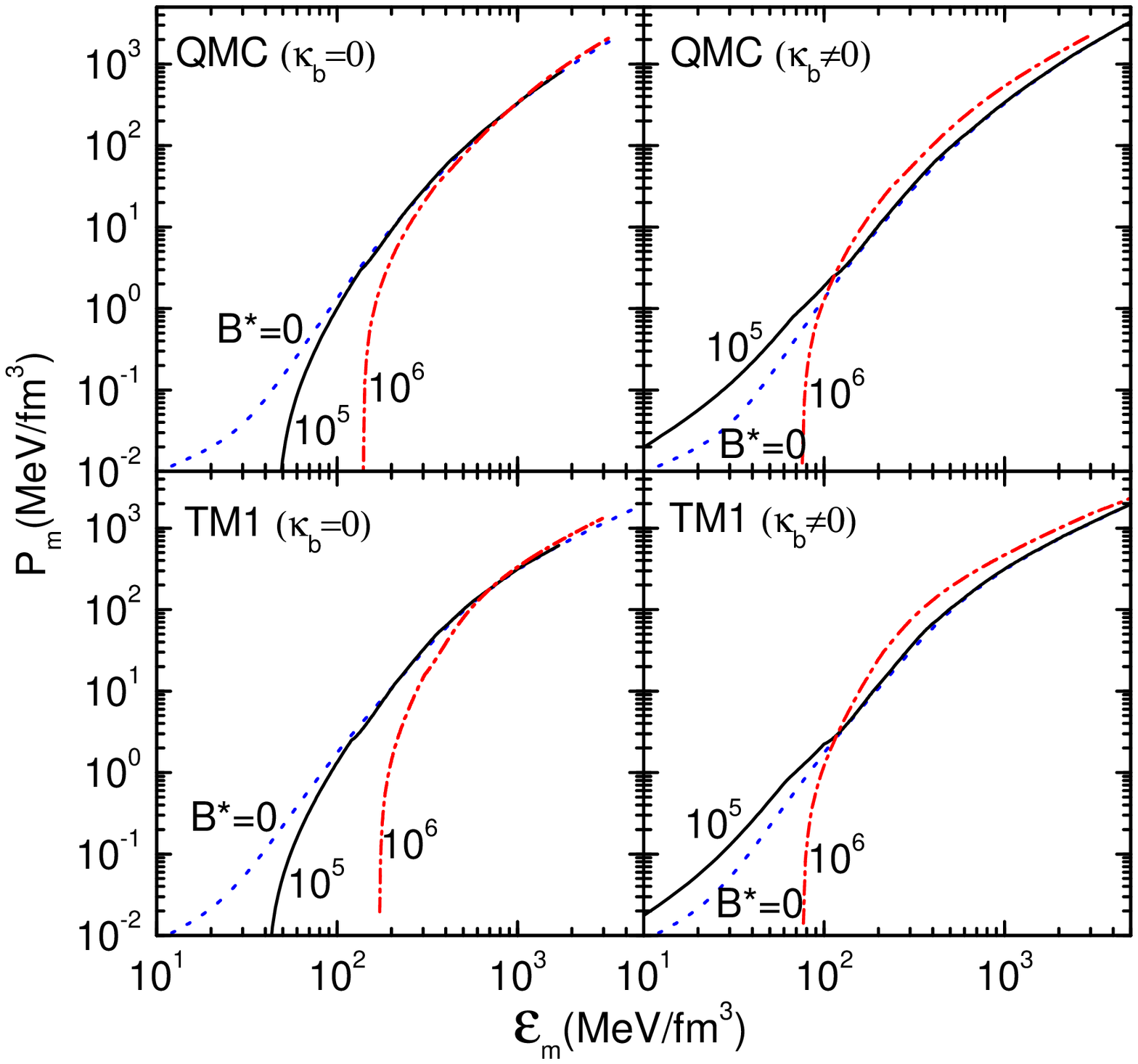}
\caption{(Color online) The matter pressure $P_m$ versus the matter energy
density $\varepsilon_m$ for the field strengths
$B^{\ast }=0$, $10^{5}$, and $10^{6}$.
The results of the QMC model with and without the inclusion of anomalous
magnetic moments are shown in the upper-right and upper-left panels,
respectively. Those in the TM1 model are also shown in the lower panels
for comparison.}
\label{fig:pe}
\end{figure}

\begin{figure}[htb]
\includegraphics[bb=30 280 535 755, width=12 cm,clip]{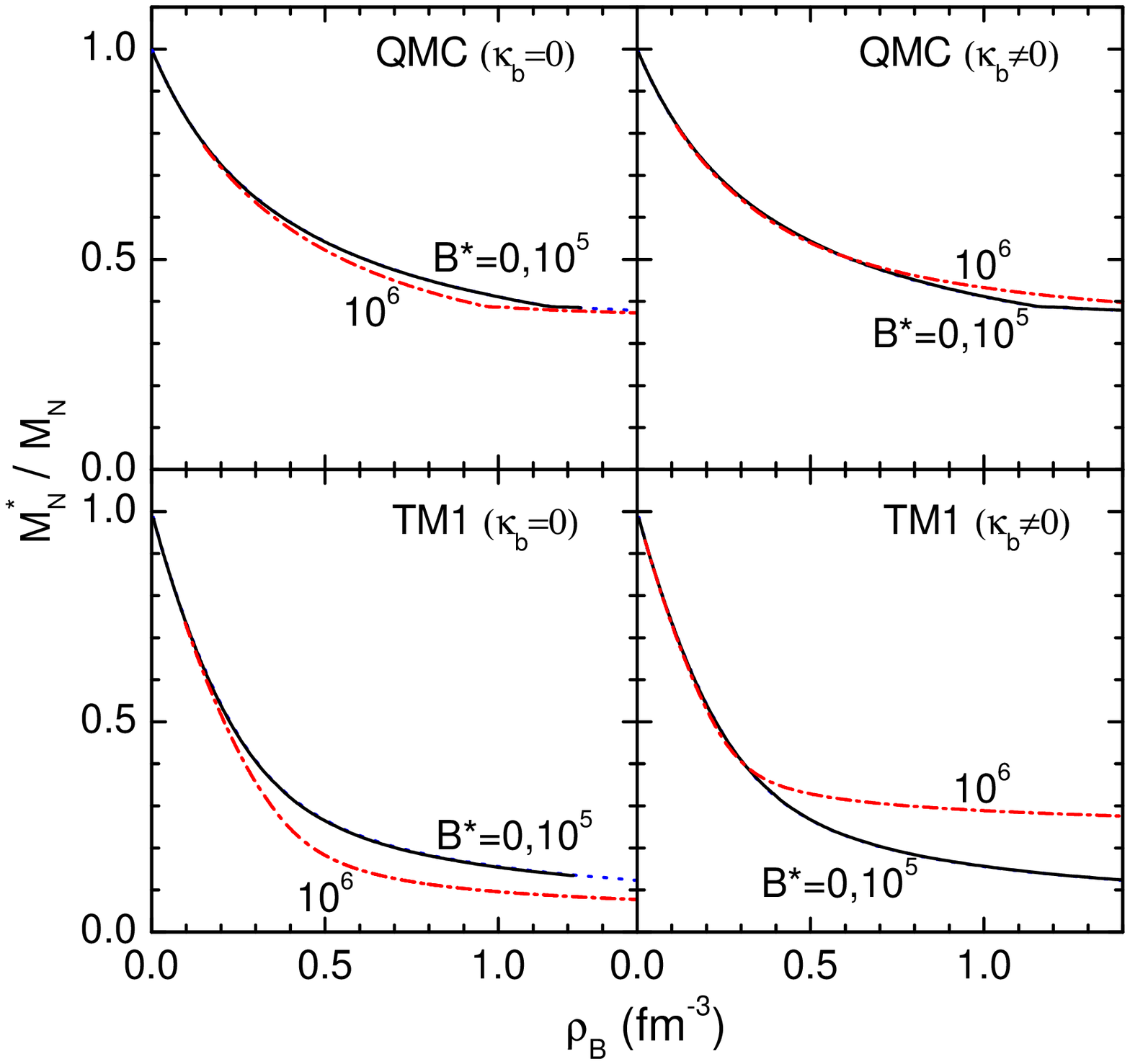}
\caption{(Color online) The effective nucleon mass $M_{N}^{\ast}/M_{N}$ as a function
of the baryon density $\rho _{B}$ for the field strengths
$B^{\ast }=0$, $10^{5}$, and $10^{6}$. The results with and without
the inclusion of anomalous magnetic moments are shown in the right
and left panels, respectively.}
\label{fig:mrho}
\end{figure}

In Fig.~\ref{fig:pe}, we show the matter pressure $P_m$ as a function of the matter
energy density $\varepsilon_m$ for the field strengths $B^{\ast }=B/B_{c}^{e}=0$,
$10^{5}$, and $10^{6}$. The results of the QMC model in the upper panels are compared
with those of the RMF model with TM1 parameter set~\cite{ST94} in the lower panels.
In Ref.~\cite{Latt00}, the RMF model with GM3~\cite{GM91} and several other parameter
sets has been used and compared with each other. Here we prefer to make a comparison
with the TM1 model, which includes nonlinear terms both for $\sigma $ and
$\omega $ mesons. The TM1 model was determined in Ref.~\cite{ST94} by reproducing
the properties of nuclear matter and finite nuclei including neutron-rich nuclei,
and has been widely used in many studies of nuclear physics~\cite{Kaon02,npa98,hs96,prc02}.
We show the results with and without the inclusion of nucleon anomalous magnetic
moments in the right and left panels, respectively.
In both QMC and TM1 models, the Landau quantization of charged
particles causes a softening in the EOS as shown in the left panels, while
the inclusion of anomalous magnetic moments leads to a stiffening of the EOS
as shown in the right panels. These effects become significant as magnetic fields
increase above $B^{\ast } \sim 10^{5}$. The softening of the EOS due to Landau
quantization can be overwhelmed by the stiffening due to the incorporation
of anomalous magnetic moments with increasing magnetic fields for $B^{\ast
}>10^{5}$. The EOS in the QMC model is slightly stiffer than the one in the
TM1 model.
We show in Fig.~\ref{fig:mrho} the effective nucleon masses as a function of
baryon density $\rho _{B}$ again for $B^{\ast }=0$, $10^{5}$, and $10^{6}$.
It is found that the influence of magnetic fields on the effective masses is
not observable until $B^{\ast }>10^{5}$. Without the inclusion of anomalous
magnetic moments, the effective masses for $B^{\ast }>10^{5}$  are smaller
than the field-free values, but they become to be larger than the field-free
values in strong magnetic fields when the anomalous magnetic moments are included.
It is clear that the changes of effective masses in the QMC model are much smaller
than those in the TM1 model.

\begin{figure}[htb]
\includegraphics[bb=50 440 560 800, width=12 cm,clip]{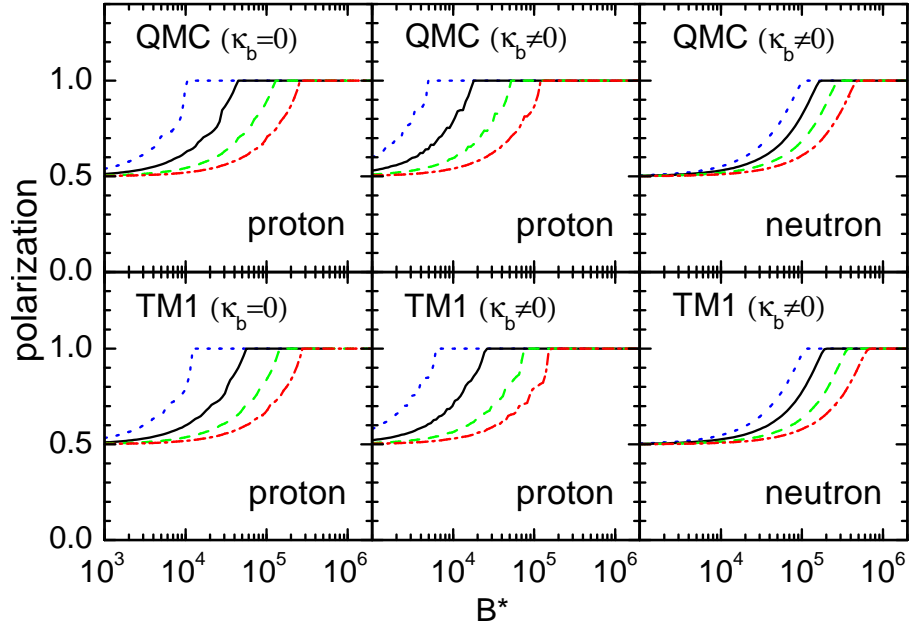}
\caption{(Color online) The spin polarization of protons (left and middle panels)
and neutrons (right panels) as functions of magnetic field strength.
The dotted (blue), solid (black), dashed (green),
and dashed-dotted (red) lines correspond to
$\rho_{B}=0.075$, $0.15$, $0.30$, and $0.60$ fm$^{-3}$, respectively.
The neutron spin polarization is purely due to the interaction of anomalous
magnetic moments with the magnetic field. The inclusion of anomalous
magnetic moments leads to an increase in the proton spin polarization.}
\label{fig:ppnb}
\end{figure}

It is instructive to discuss the spin polarization of nucleons and electrons in
neutron star matter with the presence of strong magnetic fields. For charged
particles such as electrons and protons, the energy spectrum is
characterized by the quantum number of Landau level $\nu $ and the spin
projection $s$. If we ignore the anomalous magnetic moments of the charged
particles, all energy levels except the ground state are doubly degenerate
with opposite spin projections. The ground state of electrons
(protons) is a single state with $\nu =0,$ $s=-1$ $\left( s=+1\right)$. As
the field strength increases, the energy gap between Landau levels becomes
larger and larger, and at a critical field, all particles only occupy the
first Landau level with $\nu =0,$ $s=-1$ for electrons and $s=+1$ for
protons. This is called complete spin polarization which occurs
as the energy gap between Landau levels is comparable
to the Fermi momentum of the particle. It is the order of
$B^{\ast } \sim 10^{4}$ for electrons and protons in neutron star matter at the
densities of interest. Note that neutrons can not be altered by external
magnetic fields when the anomalous magnetic moments are ignored. It is very
interesting to estimate the contributions from anomalous magnetic moments on
the spin polarization of nucleons. When the anomalous magnetic moments
are included, the double degeneracy of Landau levels for protons
disappears, and the energy spectra for both neutrons and protons are
spin-dependent. The ground state of protons (neutrons) is spin-up
(spin-down). The next level to be occupied for protons is spin-up
$\left( \nu =1,s=+1\right)$, then the third level may be spin-down
$\left( \nu =1,s=-1\right)$ or spin-up $\left( \nu =2,s=+1\right)$,
which should be determined by comparing their energy spectra. Neutrons can be
completely polarized with the inclusion of anomalous magnetic moments. The
spin-up states begin to be occupied by neutrons for $k_{f,-1}^{n2}\gtrsim
4\left\vert {\kappa _{n}}\right\vert {M_{N}^{\ast }B}$, so complete spin
polarization of neutrons appears at the order of $B^{\ast }\sim 10^{5}$.
We define the proton spin polarization as the ratio of spin-up proton density
to total proton density, $\rho _{v}^{p}(s=+1)/\rho _{v}^{p}$. The results with and
without the inclusion of anomalous magnetic moments are shown in the middle and
left panels of Fig.~\ref{fig:ppnb}, while the neutron spin polarization
defined as the ratio of spin-down neutron density to total neutron density,
$\rho_{v}^{n}(s=-1)/\rho _{v}^{n}$, is shown in the right panels. It is obvious
that both proton and neutron spin polarizations increase with decreasing baryon
density and increasing field strength, and a complete spin polarization
occurs at a critical $B^{\ast }$ for a given baryon density. By comparing the
left and middle panels, one can find that the inclusion of anomalous
magnetic moments leads to a significant increase in the proton spin polarization.
We note that there is no noticeable difference between the QMC and TM1 models
in Fig.~\ref{fig:ppnb}.

\begin{figure}[htb]
\includegraphics[bb=20 270 540 760, width=12 cm,clip]{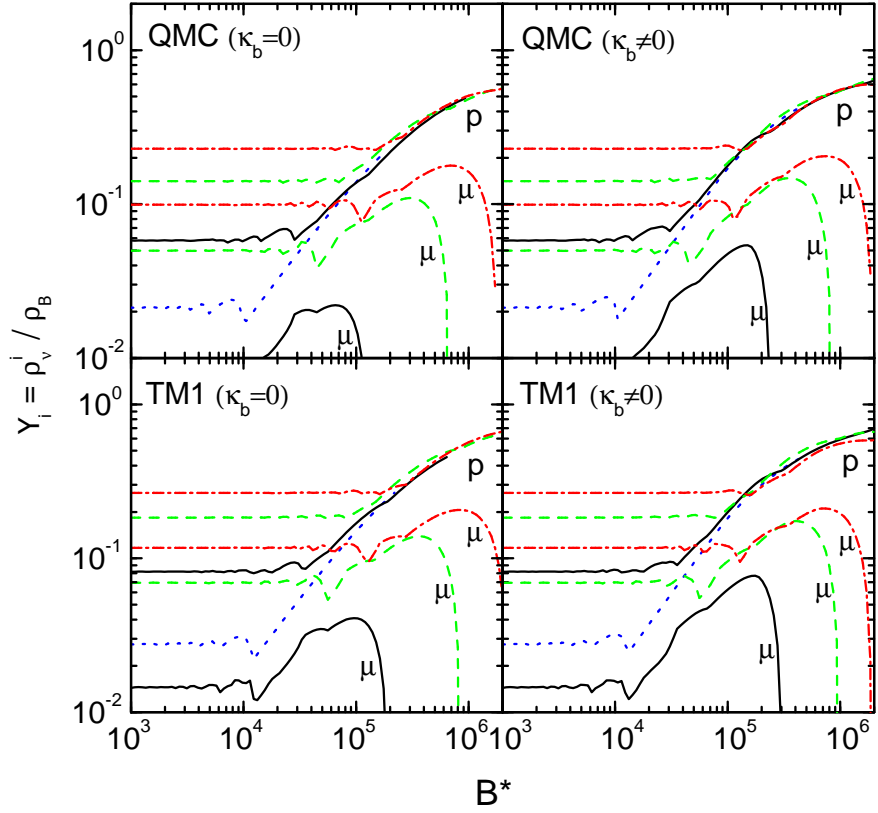}
\caption{(Color online) The proton and muon fractions as functions of
magnetic field strength. The dotted (blue), solid (black), dashed (green),
and dashed-dotted (red) lines correspond to
$\rho_{B}=0.075$, $0.15$, $0.30$, and $0.60$ fm$^{-3}$, respectively.
The neutron and electron fractions could be obtained by
$Y_{n}=1-Y_{p}$ and $Y_{e}=Y_{p}-Y_{\mu}$.}
\label{fig:yb}
\end{figure}

In Fig.~\ref{fig:yb}, the proton fraction $Y_{p}=\rho _{v}^{p}/\rho _{B}$
and the muon fraction $Y_{\mu }=\rho _{v}^{\mu }/\rho _{B}$ are plotted as
functions of the field strength $B^{\ast }$ for the baryon densities
$\rho_{B}=0.075$, $0.15$, $0.30$, and $0.60$ fm$^{-3}$. One can easily obtain
the neutron fraction $Y_{n}=\rho_{v}^{n}/\rho _{B}$ and the electron fraction
$Y_{e}=\rho _{v}^{e}/\rho _{B}$\ by the relations: $Y_{n}=1-Y_{p}$
and $Y_{e}=Y_{p}-Y_{\mu }$. The results with and without the
inclusion of anomalous magnetic moments are shown in the right and left
panels. It is obvious that the composition of neutron star matter depends on
both the baryon density $\rho_{B}$ and the magnetic field strength $B^{\ast}$.
When the external magnetic fields are absent, the composition of matter is
dominated by neutrons with a small proton fraction. It is clear that the proton
fraction remains unaltered from the field-free case for relatively low fields.
As the field strength increases to the value of having complete proton spin
polarization, a significant enhancement of the proton fraction is observed
due to large reductions in chemical potentials of protons and leptons caused
by strong magnetic fields, and then protons become to be dominant for $B^{\ast
}>10^{6}$. On the other hand, lepton fractions increase with the increasing
of proton fraction by the requirement of charge neutrality. The effect of
strong magnetic fields leads to a reduction in the electron chemical
potential, so that muons disappear when the electron chemical potential is less
than the rest mass of the muon. It is evident that the inclusion of
anomalous magnetic moments causes a slight increase in the proton and lepton
fractions.


\section{Summary}
In this paper we have investigated the effects of strong magnetic fields
on the properties of neutron star matter within the QMC model.
The QMC model describes a nuclear many-body system
as nonoverlapping MIT bags in which quarks interact through self-consistent
exchange of scalar and vector mesons in the mean-field approximation.
The effective nucleon masses in the QMC model are obtained self-consistently
at quark level, which is an important difference from the RMF models.
It is clear that the effects of strong magnetic fields become significant
only for field strength $B^{\ast}>10^{5}$. The Landau quantization of
charged particles causes a considerable enhancement of the proton fraction
and a softening in the EOS, while the inclusion of nucleon anomalous magnetic moments
leads to a stiffening of the EOS. With the inclusion of anomalous magnetic
moments, a complete spin polarization of fermions occurs at a
critical $B^{\ast}$ which depends on the baryon density. The complete spin
polarization of neutrons appears at higher field than the one of protons.

We have made a systematic comparison between the results of the QMC and RMF models.
It is found that quantitative differences exist between these models,
while qualitative trends of magnetic field effects on the EOS
and composition of neutron star matter are very similar.
In the interior of neutron stars hadronic matter may undergo
a phase transition to deconfined quark matter. To estimate the phase transition
in the presence of strong magnetic fields, the QMC model should be applicable since
it describes hadronic matter in terms of quark degrees of freedom.

\section*{Acknowledgments}

This work was supported in part by the National Natural Science Foundation
of China and the Specialized Research Fund for the Doctoral
Program of Higher Education (No. 20040055010).



\newpage

\newpage

\newpage

\newpage

\end{document}